\documentclass{svproc}
\usepackage{url,graphicx,subfig,amsfonts, amssymb,amsmath,bm,tabularx}

\begin{document}
\mainmatter              
\title{Energy dependence of the $\bar{K}N$ interaction and the two-pole structure of the $\Lambda(1405)$  -- are they real? }
\titlerunning{Energy dependence of the $\bar{K}N$ interaction}  
%
\author{J\'anos R\'evai\inst{}}
\authorrunning{J. R\'evai} 

\institute{MTA Wigner RCP, Budapest, Hungary
\\
\email{revai.janos@wigner.mta.hu}}

\maketitle              

\begin{abstract}
It is shown, that the energy-dependence of the chiral based $\bar{K}N$ potentials, responsible for the 
occurence of two poles in the $I=0$ sector is the consequence of applying the on-shell factorization 
approximation \cite{Oset_Ramos}. When the dynamical equation is solved without this approximation, 
the $T$-matrix has only one pole in the region of the $\Lambda(1405)$ resonance.
\end{abstract}
\subsection*{Introduction} 

 The $\Lambda(1405)$ is one of the basic objects of strangeness nuclear physics (SNP). Experimentally
it is a well pronounced bump in the $\pi\Sigma$ missing mass spectrum in various reactions somewhat below
the $K^-p$ threshold with PDG resonance parameters $E-i\Gamma/2=(1405-25i)MeV$.
Theoretically it is an $I=0$ quasi-bound state in the $\bar{K}N-\pi\Sigma$ system, which decays into the
$\pi\Sigma$ channel.

Constructing any multichannel $\bar{K}N$ interaction, the starting point of any SNP study, one of the first
questions is :"What kind of   $\Lambda(1405)$ it produces?" At present, it is believed, that theoretically
substantiated $\bar{K}N$ interactions can be derived from the chiral perturbation expansion of the $SU(3)$
meson-baryon Lagrangian. For these interactions the widely accepted answer to the above question is,
that the observed  $\Lambda(1405)$ is the result of the interplay of two $T$-matrix poles. Our aim is to
challenge this opinion.

\subsection*{The full and on-shell factorized WT potentials $\mathbf{ \hat{V}}$ and $\mathbf{\hat{U}}$.}
Our starting point is the lowest order Weinberg-Tomozawa (WT) term of the chiral Lagrangian ( eq.(7) from the basic paper \cite{Oset_Ramos}):
\begin {equation}
\label{orig}
\langle q_i |v_{ij}|q_j\rangle \sim - {c_{ij} \over 4 f_{\pi}^2}(q_i^0+q_j^0),
\end{equation}
where $i$ and $j$ denote the different meson-baryon channels ($i,j=1,2,3,4,5=[\bar{K}N]^{I=0},[\bar{K}N]^{I=1},[\pi\Sigma]^{I=0},
[\pi\Sigma]^{I=1},[\pi\Lambda]^{I=1}$), $q_i$ and $q_i^0=\sqrt{m_i^2+q_i^2}$ denote the meson c.m. momentum and energy,
$c_{ij}$ are $SU(3)$ Clebsch-Gordan coefficients and $f_{\pi}$ is the pion decay constant, $m_i(M_i)$ are the meson (baryon) masses.
Physical quantities can be derived from this expression via a certain dynamical framework, relativistic (BS equation, relativistic kinematics)
or non-relativistic (LS equation, non-relativistic kinematics). We shall use the second option, having  
in mind applications for $n>2$ systems. According to our choice and the usual practice, eq.(\ref{orig}) has to be supplemented: adding appropriate 
normalization factors, applying a relativistic correction to meson energies and introducing two meson decay constants instead of $f_\pi$:
 \begin{equation}
 \label{vnorm}
 \langle q_i|v_{ij}|q_j\rangle=-{c_{ij}\over{ 64 \pi^3 F_iF_j\sqrt{m_im_j}}}({q_i^0}'+{q_j^0}'),
 \end{equation}
\goodbreak
where
 \begin{equation}
  {q_i^0}'=q_i^0+{{q_i^0}^2-m_i^2\over 2 M_i}=q_i^0+{q_i^2\over 2 M_i}\,\, {\underset{\rm nonrel}\approx}\,\,m_i+{q_i^2\over 2 \mu_i}
  \end{equation}
  with the reduced mass $\mu_i=m_i M_i/(m_i+M_i)$ and $F_i$, $i=\bar{K},\pi$ are the new meson decay constants. In order to use the potential (\ref{vnorm}) 
in LS equation a regularization procedure has to be applied to ensure  convergence of the occuring integrals. We use the separable potential
 representation of the interaction, which amounts to multiplying the potential (\ref{vnorm}) by suitable cut-off factors $u_i(q_i)$ and $u_j(q_j)$.
Finally, the potential $V_{ij}$ entering the LS equation for total energy $W$ 
\begin{equation}
\label{LS}
\langle q_i|T_{ij}(W)|q_j\rangle =\langle q_i|V_{ij}|q_j\rangle+\sum_s\int \langle q_i|V_{is}|q_s\rangle G_s(q_s;W)\langle q_s|T_{sj}(W)|q_j\rangle d\vec{q}_s
\end{equation}
has the form
\begin{equation}
  \langle q_i|V_{ij}|q_j\rangle=u_i(q_i) \langle q_i|v_{ij}|q_j\rangle u_j(q_j)= \lambda_{ij}(g_{iA}(q_i)g_{jB}(q_j)+g_{iB}(q_i)g_{jA}(q_j))
 \end{equation} 
which is a two-term multichannel separable potential with form factors
$$
 g_{iA}(q_i)=u_i(q_i); \ \ g_{iB}(q_i)=g_{iA}(q_i)\gamma_i(q_i);\ \ \gamma_i(q_i)=(m_i+{q_i^2\over 2\mu_i}),\ \ 
$$
and coupling matrix
$$
 \lambda_{ij}=-{c_{ij}\over 64 \pi^3 F_iF_j\sqrt{m_im_j}}.
$$ 
The non-relativistic propagator $G_s(q_s;W)$ has the form
\begin{equation}
\label{prop}
G_s(q_s;W)=(W-m_s-M_s-{q_s^2\over 2 \mu_s}+i\epsilon)^{-1}={2\mu_s\over k_s^2-q_s^2+i\epsilon },
\end{equation}
where $k_s=\sqrt{2\mu_s (W-m_s-M_s)}$ is the on-shell c.m. momentum in channel $s$.

A commonly used procedure before solving the integral equation (\ref{LS}) is to remove the inherent $q$-dependence of the potential
by replacing $q_i$ in $\gamma_i(q_i)$ by its on-shell value $k_i$: $\gamma_i(q_i)\rightarrow \gamma_i(k_i)=W-M_i$. This is the so-called
on-shell factorization approximation, introduced in \cite{Oset_Ramos} and never checked afterwords. The separable potential representation 
of the interaction allows an exact solution of the LS equation (\ref{LS}) for both versions of the potential: the "full" WT potential
\begin{equation}
\label{V}
\hat{V}=\sum_{i,j}|g_{iA}\rangle\lambda_{ij}\langle g_{jB}|+|g_{iB}\rangle\lambda_{ij}\langle g_{jA}|
\end{equation}
and its on-shell factorized energy-dependent counterpart
\begin{equation}
\label{U}
\hat{U}(W)=\sum_{i,j}|g_{iA}\rangle\lambda_{ij}(2W-M_i-M_j)\langle g_{jA}|
\end{equation}
providing thus a check of the effects of this approximations. 
Moreover, the separable potential approach offers
an insight into the nature of the on-shell factorization approximation. When calculating
the Green's function matrix elements containing $g_{iB}$-type form-factors,  e.g.
\begin{equation*}
\langle g_{iA}|G_i(W)|g_{iB}\rangle=2\mu_i\int {u_i(q)^2\gamma_i(q)\over k_i^2-q^2+i\varepsilon}d\vec{q},
\end{equation*}
the on-shell factorization replaces $\gamma_i (q)$ under the integration sign by $\gamma_i (k_i)=W-M_i$ 
and puts it outside the integral. It can be seen, that for real, positive $k_i$, when the integral is singular this might
have some justification, however, for complex (or imaginary) $k_i$, which is the case when bound states or complex pole positions are sought,
the approximation seems to be meaningless.
\subsection*{Numerical results}
\begin{figure}[h!]
\includegraphics[width=\textwidth]{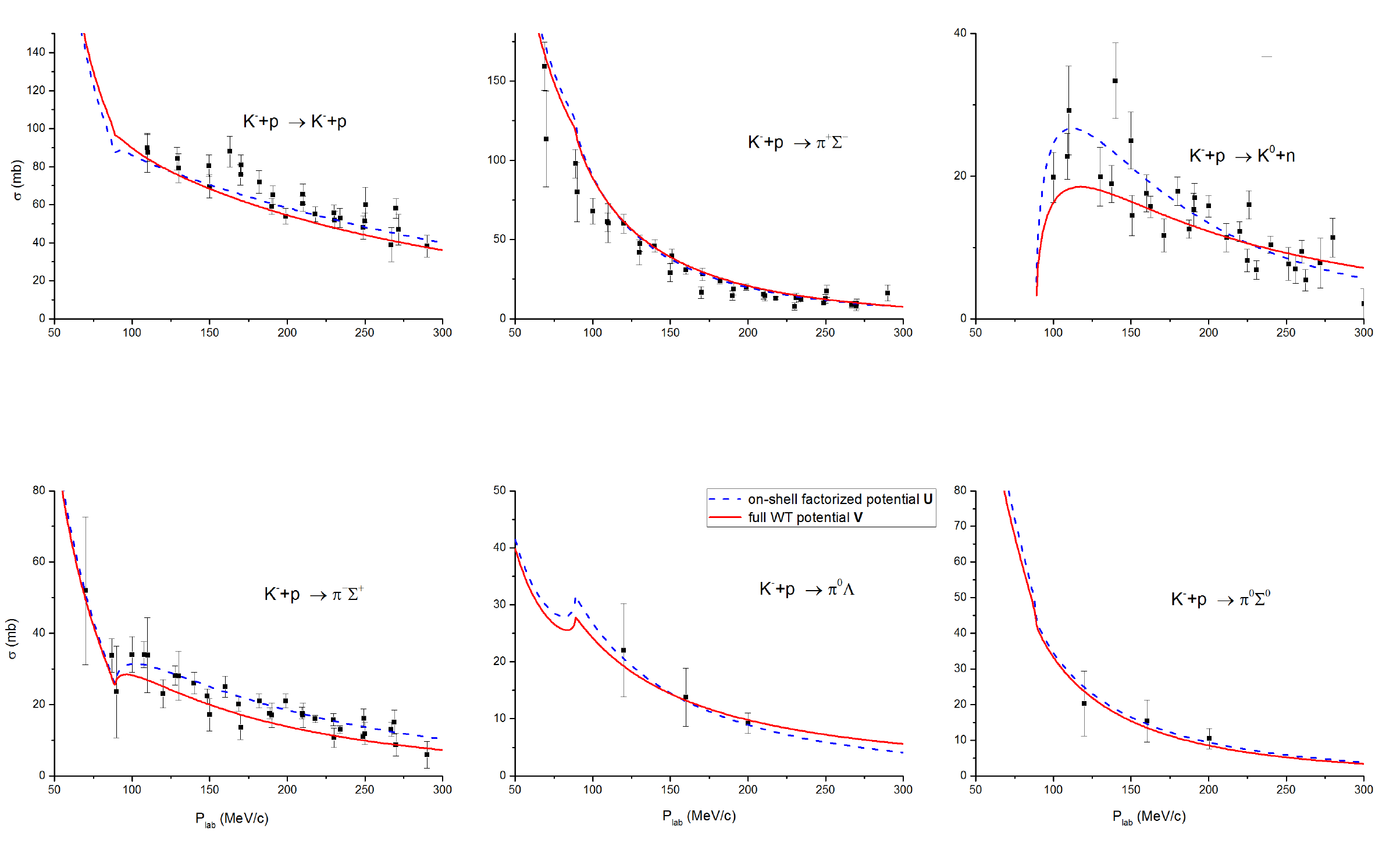}
\caption{Elastic and inelastic $K^-p$ cross sections for the potentials $\bm{U}$ and $\bm{V}$.}
\label{fig:cross}
\end{figure}

Practical solution of eq.(\ref{LS}) starts with an appropriate choice of the form- or cut-off factors $u_i(q)$, which ensures the convergence of all
occuring integrals. In our case it was the dipole Yamaguchi form with adjustable cut-off (or range) parameters $\beta_i$:
$$
u_i(q)=\left({\beta_{i}^2\over {q^2+\beta_{i}^2}}\right)^2
$$
The details of the formalism for the actual calculations can be found in \cite{revai}.  Both potentials $\hat{V}$ and $\hat{U}$ depend on the same set
of 7 parameters $F_{\bar{K}},F_{\pi},\beta_1,\beta_2,\beta_3,\beta_4$ and $\beta_5$ which have to be fitted to the available experimental data,
which are the 6 low-energy elastic and inelastic $K^-N$ cross sections, the 3 threshold branching ratios $\gamma, R_n, R_c$\footnote{For their definition
see \cite{revai}} and the $1s$ level shift $\Delta E$ in kaonic hydrogen. The results of the fit for the two potentials are shown in 
Table \ref{fit} and  Fig.\ref{fig:cross} . Table \ref{par} shows the obtained parameter values.

More or less equal quality fits can be obtained for both potentials but for very different parameter values. This means, that  $\hat{U}$ can not be considered as an approximation to $\hat{V}$ -- they are basically different interactions. Their most significant difference is, that, while the full WT potential  $\hat{V}$ produces a 
single pole in the region of $\Lambda(1405)$, the on-shell factorized potential  $\hat{U}$ for any reasonable combination of parameters produces the 
familiar two poles. The pole positions for the two potentials are
\begin{equation*}
\begin{split}
 z1&=(1425-21i) MeV\ \ {\rm for}\ \ \hat{V}\ \ {\rm  and} \\
 z1&=(1428-35i) MeV\ \  {\rm and}\ \  z2=(1384-62i) MeV\ \ {\rm for}\ \  \hat{U}. 
\end{split}
\end{equation*}
The position of the single $\hat{V}$ pole does not confirm the strong $K^-p$ binding, which is the main feature of the phenomenological potentials adjusted to the PDG pole position.

\begin{table}[h]
\begin{center}
\begin{tabular}{p{2.5cm}p{2.5cm}p{2.5cm}p{4cm}}
\hline
&$\hat{V}$&$\hat{U}$&Exp\\
\hline
$\gamma$&$2.32$&$2.35$&$2.36\pm0.04$\\
$R_c$&$0.671$&$0.664$&$0.664\pm0.011$\\
$R_n$&$0.202$&$0.194$&$0.189\pm0.015$\\
$\Delta E\ (eV)$&$350-279\ i$&$302-294\ i$&$(283\pm36)-(271\pm46)\ i$\\
\hline
\end{tabular}
\end{center}
\caption{Calculated and experimental values of the discrete data for potentials  $\hat{U}$ and $\hat{V}$\vspace{-.3cm}.}
\label{fit}
\end{table}

\begin{table}[h]
\begin{center}
\begin{tabular}{p{1.4cm}p{1.4cm}p{1.4cm}p{1.4cm}p{1.4cm}p{1.4cm}p{1.4cm}p{1.4cm}}
\hline
&$F_{\pi}$&$F_K$&$\beta_1$&$\beta_2$&$\beta_3$&$\beta_4$&$\beta_5$\\
\hline
$\hat{V}$&80.8&132&1094&960&516&537&629\\
$\hat{U}$&107&109&1247&1622&919&959&443\\
\hline
\end{tabular}
\end{center}
\caption{Parameters of the potentials $\hat{V}$ and $\hat{U}$ (all values are given in $MeV$\vspace{-.5cm}).}
\label{par}
\end{table}
Most recent and accurate information on the $\Lambda(1405)$  resonance comes from the CLAS photoproduction experiment 
 $\gamma + p \rightarrow K^+ +\pi^0+\Sigma^0$ \cite{clas} in which the $M(\pi^0\Sigma^0)$ missing mass spectra give
 the $\Lambda(1405)$ line shape. For the analysis of these spectra at present we have the semi-phenomenological final state
interaction formula of Roca and Oset \cite{Roca_Oset}, which contains some adjustable parameters $c_i$ and the $T$-matrix
elements of the $\bar{K}N-\pi\Sigma-\pi\Lambda$ potential. Using energy-dependent (two-pole) $\bar{K}N$ potentials, with 
simultaneous variation of $c_i$ and the potential parameters, acceptable fits to the CLAS data were obtained. This fact then was 
considered as the ultimate proof of the two-pole structure of   the $\Lambda(1405)$.

\begin{figure}
\includegraphics[width=\textwidth]{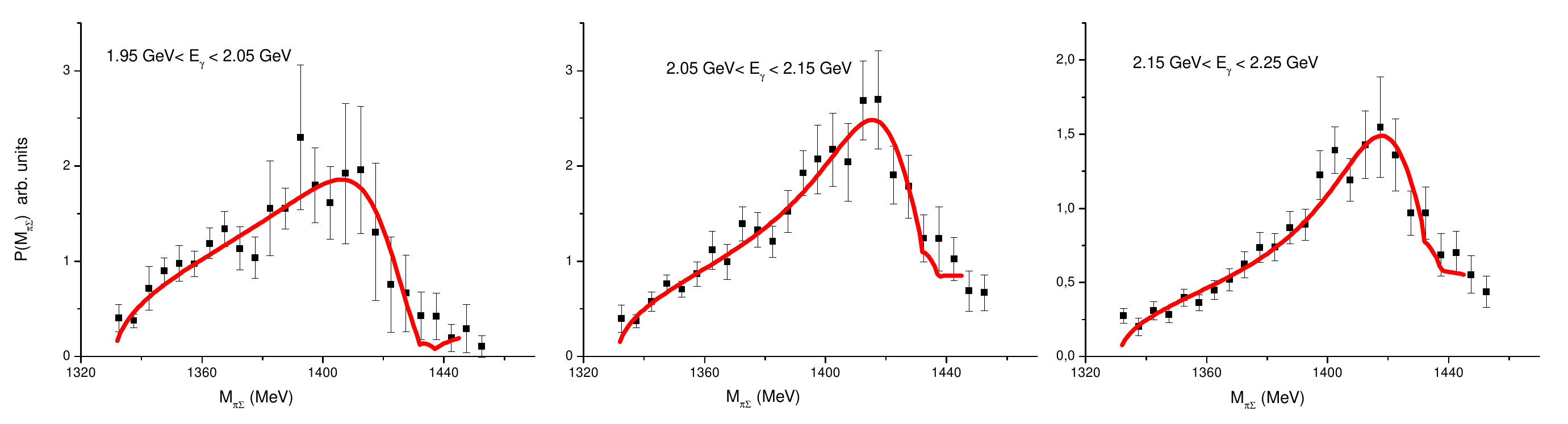}
\caption{Fit to the first 3 $\gamma$-energy bins of CLAS $M(\pi^0\Sigma^0)$ missing mass spectra in
$\gamma + p \rightarrow K^++\pi^0+\Sigma^0$ photoproduction.}
\label{fig:clasfit}
\end{figure}

Using the same formula (corrected for the repeated use of on-shell factorization), we have made a preliminary fit to the few lowest 
$\gamma$-energy bin CLAS data varying only the $c_i$-s with our unchanged single-pole potential $\hat{V}$. The results are shown in 
Fig.\ref{fig:clasfit}.

It does not seem, that another pole is necessarily needed to improve these fits. A complete analysis of the CLAS data, including
the charged channels is the subject of a forthcoming work.
\subsection*{Conclusions}
\begin{itemize}
\item
It was shown, that the energy-dependence of the WT term of the $\bar{K}N$ interaction, derived from the 
chiral $SU(3)$ Lagrangian and responsible for the appearance of a second pole in the $\Lambda(1405)$ region, 
follows from the unjustified application of the on-shell factorization approximation.
\item
Without this approximation a new, chiral based, energy-independent $\bar{K}N$ interaction was derived, which 
supports only one pole in the region of the   $\Lambda(1405)$ resonance.
\item
The widely accepted "two-pole structure" of the $\Lambda(1405)$ state thus becomes questionable.
\item
In coordinate space calculations for $n>2$ systems the use of the new potential avoids the not easily (and not uniquely)
surmountable difficulties arising from the energy dependence of the two-body interaction.
\end{itemize}
The work was supported by the Hungarian OTKA grant  109462.

%
%

\end{document}